**Bennett, CC. (2012). Utilizing RxNorm to Support Practical Computing Applications: Capturing Medication History in Live Electronic Health Records.** *Journal of Biomedical Informatics*. **45: 634-641.** http://www.sciencedirect.com/science/article/pii/S153204641200041X

# Utilizing RxNorm to Support Practical Computing Applications: Capturing Medication History in Live Electronic Health Records

Casey C. Bennett[a,b]

[a]*Dept. of Informatics*
*Centerstone Research Institute*
*Nashville, TN, USA*

[b]*School of Informatics and Computing*
*Indiana University*
*Bloomington, IN, USA*

Corresponding Author:
*Casey Bennett*
*Department of Informatics*
*Centerstone Research Institute*
*365 South Park Ridge Road*
*Bloomington, IN 47401*
1.812.337.2302
Casey.Bennett@CenterstoneResearch.org

**Abstract**

RxNorm was utilized as the basis for direct-capture of medication history data in a live EHR system deployed in a large, multi-state outpatient behavioral healthcare provider in the United States serving over 75,000 distinct patients each year across 130 clinical locations. This tool incorporated auto-complete search functionality for medications and proper dosage identification assistance. The overarching goal was to understand if and how standardized terminologies like RxNorm can be used to support practical computing applications in live EHR systems. We describe the stages of implementation, approaches used to adapt RxNorm's data structure for the intended EHR application, and the challenges faced. We evaluate the implementation using a four-factor framework addressing flexibility, speed, data integrity, and medication coverage. RxNorm proved to be functional for the intended application, given appropriate adaptations to address high-speed input/output (I/O) requirements of a live EHR and the flexibility required for data entry in multiple potential clinical scenarios. Future research around search optimization for medication entry, user profiling, and linking RxNorm to drug classification schemes holds great potential for improving the user experience and utility of medication data in EHRs.

# 1. Introduction

RxNorm is a standardized, controlled terminology for medications in the United States. It includes multiple components – medication name (both generic and brand), dosage, route of administration, ingredients, and fully-specified "common dose forms" (i.e., what a physician might enter as part of a prescription to a pharmacy) [1]. These multiple components are linked together through a relational file structure, easily portable into database format [2]. RxNorm was originally developed as part of the Unified Medical Language System (UMLS) effort to integrate and map diverse and competing controlled medical terminologies in order to facilitate interoperability and data exchange across healthcare providers [3]. Early work that led to the development of RxNorm in UMLS was spurred by efforts of the HL7 Vocabulary Technical Committee in the late 1990's [1].

A major goal in developing RxNorm was to provide a consistent, standardized way to identify the essential components of prescriptions for purposes of facilitating electronic capture of such data in electronic health records (EHRs) to support the exchange of patient information across providers (i.e., interoperability), development of clinical decision support systems (CDSS), quality improvement efforts, and other research endeavors. Early efforts to utilize RxNorm have primarily focused on: 1) navigating between names/codes across differing vocabularies and home-grown legacy systems, 2) exchanging data across providers, and 3) developing medication-related CDSS [1]. Many of these efforts have taken the approach of mapping existent EHR data into RxNorm format post-hoc [2,4], although there are an increasing number of efforts to develop "direct capture" methods where prescription data are captured in RxNorm format at the point of data entry [1,2,5]. However, there is limited evidence about the use and implementation of RxNorm in actual clinical practice outside of such research settings [2].

RxNorm has taken on increased importance with development of Meaningful Use standards in the United States, which specified the use of RxNorm for medication data in early drafts of the standards [6]. Although the explicit requirement was removed in the final draft for Stage 1 Meaningful Use, the standards still encourage RxNorm use, and there is a strong belief that RxNorm or an equivalent tool will be required at Stage 2 or 3 of Meaningful Use when those standards are released in order to achieve the overarching goal of interoperable health records.

As such, an increasing need exists to look for alternate methods (averse to, e.g., post-hoc mapping) for capturing and transforming EHR medication data into RxNorm format. This includes new e-prescriptions as they are ordered (i.e., e-prescribing), as well as medications previously prescribed (i.e., medication history) [7]. Medication history in a clinical (not pharmacy) context includes such history (at least within some reasonable time frame, e.g., the past two years) for a patient *prior* to the current provider's treatment episode, often provided by the patient themselves. Collection of such information is specified as part of Meaningful Use, and such data can be critically important for identification of proper treatment, adjusting baseline outcomes, and reducing medical errors due to drug interactions [8]. For example, a patient suffering from Major Depression who was on citalopram 20 mg until 3 months prior to the start of treatment with a new provider may not represent a true baseline due to residual effects of the previous treatment [9,10]. In order to evaluate such a patient, as well as to properly account for potential outcome improvement (or lack thereof), it is thus essential to know the patient's medication history. This sort of baseline adjustment would also be a necessity in any pay-for-performance paradigm.

In the future, patient medication history information will theoretically be available electronically via data exchange across provider EHRs. However, such electronic exchange is not imminent, and current methods for collecting medication history present broad challenges [1]. For the time being, systems are needed that are capable of capturing such history in a standardized format in the current provider's EHR. Such systems would also establish the necessary data infrastructure for future electronic exchange of standardized medication history data across providers, once such capabilities become widely available. Additionally, since medication history is typically captured via patient self-report, techniques for applying RxNorm for capturing medication history in live clinical settings and EHRs are directly applicable to capturing similar information in research study settings [2].

There are two possible methods for capturing medication history, as mentioned above: 1) Mapping data from legacy systems into RxNorm post-hoc, or 2) Direct capture of data in RxNorm-compatible format at the point of data entry. Challenges with the former scenario in real-world EHR systems include the fact that only a certain percentage of medications map correctly, even with sophisticated algorithms [11] or intensive, post-hoc manual matching by researchers, and that many successful mapping efforts have been limited to well-defined subsets containing only fully-specified prescriptions incorporating National Drug Codes (NDCs) [12]. Even the most sophisticated algorithms cannot overcome all the limitations of poorly collected medication history at the point-of-care, such as open-ended free text fields, non-standard abbreviations, or invalid combinations due to uncontrolled data capture (e.g., recording a medication in a non-existent dose unit or dose form). Limiting medication data to particular subsets or engaging in post-hoc manual mapping efforts is not feasible for real-world systems. Direct capture avoids some of these issues; however, there are challenges with this approach as well. For example, a medication missing in a search list can prevent data entry from occurring (i.e., "medication coverage"). The direct capture approach also necessitates transforming the complex relational structure of the original RxNorm tables into a bare-bones structure to populate fields/search boxes in a web screen that can support the high-speed input/output (I/O) required for such applications, maintain high data quality, and allow for flexibility in data capture [2,5].  Additionally, any direct capture system must be capable of handling multiple potential clinical scenarios, such as allowing for partial data capture when the patient does not recall specific information or lacks the faculties to report complete information (e.g., an individual experiencing psychotic schizophrenia).  Search functionality used in direct capture approaches can also open up further issues around search optimization and information retrieval filtering, commonly applied to web search (see Discussion).

Here, we describe an approach to direct capture utilizing RxNorm in a live EHR system deployed in a large, multi-state outpatient behavioral healthcare provider in the United States (Centerstone), serving over 75,000 distinct patients each year across 130 clinical locations.  The overarching goal is to understand if and how standardized terminologies like RxNorm can be used to support such practical computing applications in live EHR systems.

## 2. Methods

RxNorm data was obtained from the NIH/NLM's UMLS website as part of the comprehensive UMLS download (http://www.nlm.nih.gov/research/umls/) and loaded into Centerstone's existent data warehouse infrastructure, running on Postgres 9.0, via the open-source Extraction Transformation Load (ETL) tool Pentaho Kettle (version 3.2, http://kettle.pentaho.com/).  The original version used in development was 2010AB, although 2011AA is currently being used.  Extracts of RxNorm data were made from this UMLS dataset and loaded into Centerstone's EHR production database, running on Oracle 10.2.  This EHR is a heavily customized version of the Qualifacts CareLogic system (http://www.qualifacts.com/).  The subsequent activities were performed during a complete re-design of the intake module in the EHR.  Although initial development was performed using RxNorm extracts from the UMLS dataset, equivalent (and more frequently updated – weekly rather bi-annually) RxNorm-specific download files can be obtained from the UMLS website (http://www.nlm.nih.gov/research/umls/rxnorm/docs/rxnormfiles.html).  For live clinical use, the direct monthly/weekly RxNorm files would be preferred.  For all practical purposes, the extraction of data from either source is equivalent, except for table names and some slight variation in a couple of column names (CUI >> RXCUI and AUI>>RXAUI), and subsequent technical details below can be substituted for one source or the other (code for both is available in the appendix).  It should be noted that some of the data described below (e.g., dose frequency) was derived from other terminologies such as SNOMED CT where RxNorm was lacking.

The nature of the legacy medication history data in the system, as well as new data captured directly from patients, is problematic in terms of quality and consistency.  Patients often cannot remember the exact dosage taken, let alone other aspects of prescriptions that RxNorm dictates (such as dose units, dose form, and route of administration).  Additionally, other requirements (e.g., state, payer, accreditation bodies) require only medication name and dosage (amount and units).  Meaningful Use itself does not currently specify the medication components to be captured in medication history.  Given these conditions,  for purposes of this study and the re-design process, we focused only on capturing *Clinical Drug* data (as defined by Richesson et al. [2]) – medication name and dosage (amount and units), segmenting each RxNorm component into independent fields and leaving dose form/route (e.g., tablet, capsule, topical cream) for future work.  Henceforth, we use *dosage* to refer to the combination of dose amount and dose units and *common form* to refer to the RxNorm specified components of *generic ingredient – dose amount – dose units – branded name[if applicable]*  for each commonly prescribed dosage for each medication.  Dose frequency (e.g., q.d., b.i.d.) was also captured independently of this RxNorm construct using SNOMED CT codes derived from UMLS.  We extracted SNOMED terms for the most commonly used ranges (as agreed upon separately by a Health Information Exchange Centerstone is part of comprising six of the largest outpatient behavioral healthcare providers in the United States).  This included terms such as qd – "once a day", qam – "Once a day, in the morning", qpm – "Once a day, in the evening", "qhs - Once a day, before bed", bid – "Twice Daily", tid – "Three times daily", "qid - Four times daily", qod – "Every other day," prn – "As Required", and mdu – "As Directed".   This list could be amended as needed to cover additional terms, including terms not covered by any terminology in UMLS.

**2.1 Direct Capture – Technical Implementation**

The aim of direct capture, at the suggestion of active clinicians, was to emulate the functionality of the Medicare Part D website formulary search (http://plancompare.medicare.gov/pfdn/FormularyFinder/).  This website allows active searching for medication names, an auto-complete suggestion feature as the user types (familiar to many readers from Google search), brand/generic identification, and modal popup functionality with common dosage information specific to each selected medication (to facilitate accurate dosing information capture).  The structure of RxNorm in its UMLS download structure is not directly amenable to use as the basis for such a system [5], in particular with consideration for I/O performance in a large, high-volume EHR.  For such an application to function effectively without noticeable lag-time, it is imperative to cache as much data on the server-side web application.  With the addition of the dynamically defined lists and search boxes, this means that minimizing the number and time of round-trips to the database over the network (the "I/O" in this case) is of principal concern.  This entails a strategy of normalizing the data into the smallest possible subsets that could be searched quickly via efficient indexes and used to construct the minimal web caches in order to support front-end functionality with minimal lag time, even in the face of potential hundreds or thousands of concurrent users.  In other words, we operationalized the basic premise behind modern relational databases [13].  The need thus existed for several different types of information:

1) A concise list of medication names for populating the auto-complete search feature
2) Commonly prescribed dosages displayed as the common forms for each medication (for populating the modal popup and auto-populating the entire screen, see figure 3 below and the Medicare D website above as an example)
3) A concise list of all known dose units (e.g., MG, MG/ML, MEQ/ML) to correctly delimit the selectable dose units for each medication (i.e. this prevents clinicians from selecting non-existent dose units for a particular medication)

Of particular interest was separating dose amount and dose units into separate, unlinked fields; such a design allows clinicians to manually edit either field independently to enter an atypical dosage (e.g.  off-label use).  This separation is useful for evaluating specific changes in dosage for research or analytical purposes, as well as algorithm development for CDSS purposes.  Otherwise we are comparing string fields of dosage information for variation or to identify medication changes, which can be imprecise and difficult to parse across patients.  Data integrity can be maintained by delimiting the possible dose units that a clinician can enter for a given medication to those units known to be applicable (derived from RxNorm itself).

This approach for implementation and visual display of RxNorm information is similar yet different from what has been proposed elsewhere [2,5] particularly in regards to utilizing both re-structured production tables of RxNorm data and optimized search functionality.  In this case, we were essentially re-structuring RxNorm data into specialized production tables to allow for improvements in speed and flexibility.   This is in essence the same as the general design differences necessitated in a transactional (Online Transaction Processing - OLTP) production system relative to an Online Analytical Processing (OLAP) system, such as a data warehouse.

A mockup of the screen was designed as a starting point in Balsamiq (http://balsamiq.com/, Figure 1).

Figure 1: Intake Medication History Mockup

Data was extracted from the original RxNorm UMLS tables to populate tables underlying this screen and the embedded search features. Common dosage prescriptions (e.g., "aripiprazole 10 MG [Abilify]") were pulled as the full name from the UMLS concept table (MRCONSO). The concise medication name (e.g., "Abilify") was pulled from the UMLS relationship table (MRREL) using *rela='has_ingredient'*. Dose amount and units was pulled from the UMLS attribute table (MRSAT) using *atn='RXN_STRENGTH'*. CUI and AUI unique identifiers were pulled for each medication. The AUI identifier of the generic form of brand-name medications was also pulled from the UMLS relationship (MRREL) table using *rela='tradename_of'*, except in the case of multiple generics or multiple ingredients

(as in some branded OTC drugs).  Finally, a flag was set using the "tty" field from the UMLS concept table (MRCONSO) to distinguish between generic and brand-name medications.  "Brand of" medication records were also suppressed, as suggested in previous literature [5].  The full multi-stage SQL code and DDL code for extracting this information is available in the appendix, for both the UMLS and monthly/weekly RxNorm files.  The SQL code was executed via the open-source ETL tool Kettle (see above).  There was some limited suppression of obvious exceptions, such as non-human medications (e.g., pet shampoos), which can be seen in the appendix code.

The final table structure was comprised of three tables (Med_List, Med_List_Common, and Med_List_Dose), populated by embedding the SQL code in the appendix into an ETL process (via Kettle). The search functionality utilizes the AutoComplete extender from the Ajax Control Toolkit for ASP.NET, set up to suggest possible complete medication names once the user has entered at least two letters, displaying only the top twelve results.  String matching is unanchored and can match any segment of the medication name, including beginning, middle, or end.  The end user can continue typing to refine the search, with the results automatically filtered/updated with each keystroke.  Medication Name information was cached on the server side with a 2 hour expiration policy (requiring cached data to be deleted or updated every two hours).  A 70 millisecond delay was set on the search box text for updating results, which allows the results to filter seamlessly as the user types.

**2.2 Implementation Evaluation**

In order to evaluate the success of using RxNorm as the basis for a direct-capture medication history tool, as well as to understand if and how it could be used as the underlying data structure for such practical computing applications, we propose a four-factor framework of analysis.  These factors are:

1) Flexibility – RxNorm data and the resulting applications should allow for the variety of scenarios that may play out in real-world data capture situations.
2) Speed – The application response time should be fast enough to generally be imperceptible to the end user.
3) Data Integrity – RxNorm data should provide high-quality, reliable data with consistent mappings (from the EHR data) back to the original source (RxNorm/UMLS), even in the face of flexibility (#1) needs.
4) Coverage – RxNorm should cover all commonly prescribed medications.

All of these factors are discussed in more detail in the Results. However, it should be noted that speed – measured as response time – was estimated by running the application using different medications with different medication component values (number of dose forms, number of dose units, etc.). Higher values presented a larger search space and thus longer run-times.

The above framework would be extensible to any standardized terminology implemented as part of a practical computing application in some provider EHR system. Generally, the framework would apply across most settings, though medications of interest may differ (e.g., in this study, in a behavioral healthcare setting, we are primarily interested in coverage of psychotropic medications and medications for common chronic health disorders, such as diabetes and hypertension) [2]. It is important to note that the four factors may be ascertained independently (e.g., demonstrating data integrity only), though such an approach is not typical in many real-world implementations. In such settings, the factors may compete (e.g., flexibility vs. data integrity), and the principle goal is to evaluate the factors as a whole, not individually.

The above factors are very similar to frameworks described elsewhere [1,5, 14] although the terminology and the definitions vary slightly. For instance, the framework concepts depicted in Fung et al. [5] map roughly as: unambiguity to data integrity, naturalness to flexibility, efficiency to speed, and coverage to coverage. Thus, there seems to be some emerging consensus on the approach to evaluating the implementation of standardized terminologies such as RxNorm.

**3. Results**

The final result was searchable information for 18708 known medications and over 42,000 common dose forms thereof. This information allowed clinicians (and patients) to search for previously prescribed medications and assist with capturing accurate dosing information (as seen in figures 2 and 3). The response time on the medication auto-complete search feature was measurable in milliseconds and generally imperceptible to the end user during human factors testing (described in forthcoming publications).

Figure 2: Medication Name Search

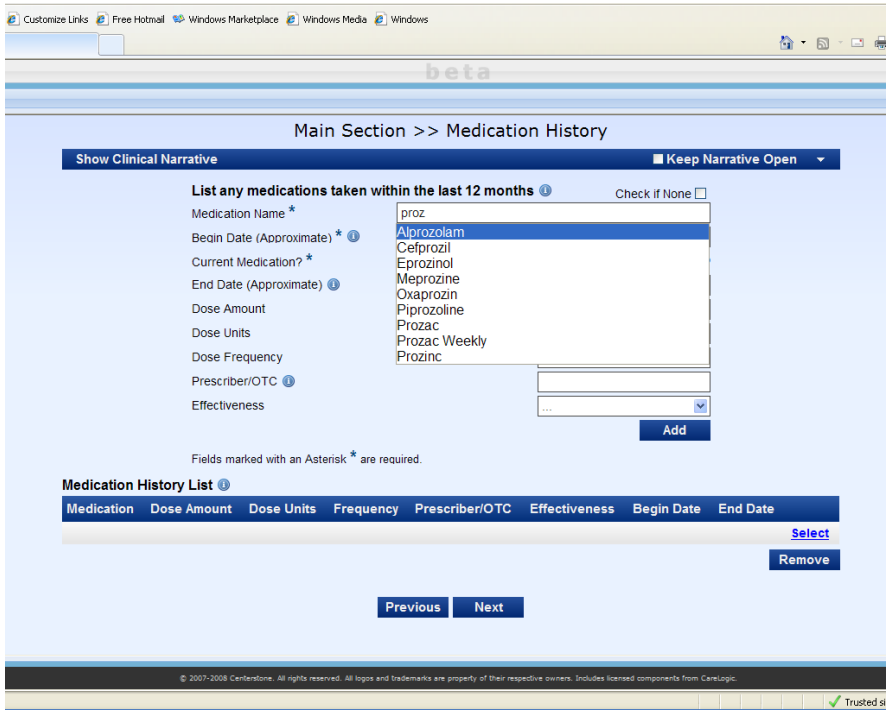

Figure 3: Medication Dosage Selector – Modal Popup

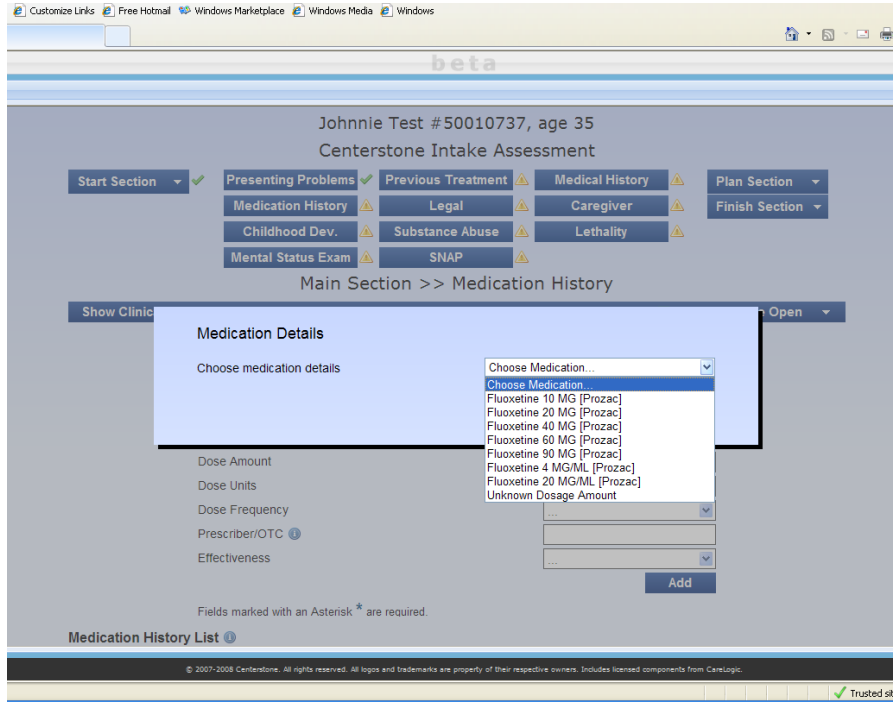

Figures 2 and 3 are driven off of 3 queries (in these examples searching for the brand name drug Prozac), with information partially cached on the server-side during end user interaction to enable efficient search/auto-complete functionality (see Methods). These queries are as follows:

**To populate the search box:**
Select med_list_id, med_name
from cen.med_list m
order by upper(med_name)
;

With "med_name" shown in the search box (See Figure 2). Selecting a particular medication opens a modal popup (Figure 3), which is populated by the following:

**To populate the modal popup:**
Select c.med_list_id, med_name, common_form, dose_amt, dose_units
from cen.med_list_common c, cen.med_list m
where c.med_list_id=m.med_list_id
and c.med_list_id=?
order by upper(med_name), dose_units, dose_amt
;

Where the question mark above is replaced by the actual med_list_id of the medication selected in the search box. The "common_form" field is shown in the popup. Once the user makes a selection (including a dummy option for "unknown"), med_name, dose_amt, and dose_units are loaded into the appropriate fields. The user can still manually adjust the dosing information at that point, which necessitates delimiting the possible dose_units for the selected medication for that dropdown field.

**To limit the options in the dose units field:**
Select med_list_id, dose_units
from cen.med_list_dose
where med_list_id=?
;

The results of the modal popup query can be seen in Table 1 (for Abilify) and Table 2 (for Zoloft). Note the query returned both the "common_form" field for selection display as well as fields for populating individual fields (e.g., dose_amt and dose_units) and returning a unique key for populating the backend tables after a selection was made.

Table 1: Modal Popup – Behind-the-Scenes Query Results - Abilify

| MED_LIST_ID | MED_NAME | COMMON_FORM | DOSE_AMT | DOSE_UNITS |
|---|---|---|---|---|
| 18949 | Abilify | aripiprazole 2 MG [Abilify] | 2 | MG |
| 18949 | Abilify | aripiprazole 5 MG [Abilify] | 5 | MG |
| 18949 | Abilify | aripiprazole 10 MG [Abilify] | 10 | MG |
| 18949 | Abilify | aripiprazole 15 MG [Abilify] | 15 | MG |
| 18949 | Abilify | aripiprazole 20 MG [Abilify] | 20 | MG |
| 18949 | Abilify | aripiprazole 30 MG [Abilify] | 30 | MG |
| 18949 | Abilify | aripiprazole 1 MG/ML [Abilify] | 1 | MG/ML |
| 18949 | Abilify | aripiprazole 7.5 MG/ML [Abilify] | 7.5 | MG/ML |

Table 2: Modal Popup – Behind-the-Scenes Query Results – Zoloft

| MED_LIST_ID | MED_NAME | COMMON_FORM | DOSE_AMT | DOSE_UNITS |
|---|---|---|---|---|
| 37694 | Zoloft | Sertraline 25 MG [Zoloft] | 25 | MG |
| 37694 | Zoloft | Sertraline 50 MG [Zoloft] | 50 | MG |
| 37694 | Zoloft | Sertraline 100 MG [Zoloft] | 100 | MG |
| 37694 | Zoloft | Sertraline 20 MG/ML [Zoloft] | 20 | MG/ML |

Selection of a common dosage, as seen in Figure 3, would then populate the remaining appropriate fields of the overall screen (medication, dosage). The end user can still manually edit any field, and an option always exists in the modal popup if the exact dosage is unknown. Additionally, an "Other Units" dummy option is provided for all medications for situations when the standard dose units are not inclusive of some atypical dosing regimen, which can occur with some less commonly used medications. Generally, such an approach allows for flexibility, at least in terms of dosing. The trade-off is that the quality of the data can be compromised when coding back into standard RxNorm common dose forms due to missing or non-standard data. However, the capability exists to reconstruct RxNorm format from the component parts via SQL query, and at a minimum the medication name can still be mapped using the CUI/AUI identifiers. Importantly, these identifiers are stored and associated with specific, controlled terms at different levels of the hierarchy, so it does not involve evaluating text after the fact. *The data are mapped as soon as collected.* In other words, even if some elements of the medication data are omitted, we can *always* map to some level of the original source (RxNorm/UMLS). The key point is that the data are always interoperable, even when collected at varying levels of detail (or if some details are omitted).

In the end, such a balance was deemed necessary to provide end users with the flexibility needed to capture medication history data across all possible clinical scenarios while still maintaining reasonable data integrity. One takeaway is that capturing medication data in fully-specified RxNorm form 100% of the time may not be a realistic goal, but capturing components of RxNorm that can be used to reconstruct an approximation of RxNorm format is achievable. As such, the system design was able to maximize data integrity without loss of flexibility.

Evaluating speed and performance is difficult, in that it is very much context-specific, particularly when dealing with networked applications. A more generalizable approach is to evaluate such a system based on principles of information theory and the branching factor of the search space [15,16]. In this case, the dynamic caching of data on the front end application reduced the number of round trips to the database from the application to a bounded upper limit of $2+(n-1)$, where n = number of medications to be entered. As the search cache is stored, we are guaranteed to hit the database at least twice (once to create the search cache, and once more to update medication-specific caches), but for subsequent medications we only need a single round trip to the database. Each round trip was estimated to average 200 milliseconds, depending on the size of the search space of the query, maxing out at around 500 milliseconds. More importantly, the cache size is reduced, whereas there are a total of over 42,000 different combinations of medication and common dosages in RxNorm. Separating out the medications into a med_list table for the initial search cache reduces this to 18,694 distinct items. The effective branching factor for common dose forms and dose units after initial medication search are 2.52 and 1.18, respectively. The represents a 55% reduction in overall cache size on average in the worst case scenario, or, assuming the caches are separated out but populated via queries, a 55% reduction in the search space to identify all medication components. In other words, given that most medications only have 2-3 common dose forms and typically just one dose unit (e.g., mg), it us much more efficient to have separate lookup tables for those distinct from the main medication list itself. This also enhances the efficiency of the keystroke-filtering auto-complete feature in the medication search box (see Methods). In both cases, the information content is still the same. Given Shannon's equation [17]:

$$H(X) = \sum_{i=1}^{n} p(x_i)I(x_i) = -\sum_{i=1}^{n} p(x_i)\log_b p(x_i)$$

Where b is typically 2. We know the *p(x$_i$)* to be 1/(total number of different combinations) for the general, non-reformatted case of RxNorm, and for the reformatted case the joint entropy *p(x$_i$)* to be the multiplicative of 1/(search size), 1/(effective branching factor of common dose form), and 1/(effective branching factor of dose units). Totaling *H(x)* for the various probabilities in each case gives us approximately ~15 bits in both cases (as expected theoretically). In other words, we reduce the needed cache size/search space and provide a bounded limit on the number of round trips to the database without loss of information. In plain language, for such an application, restructuring RxNorm can convey the same information content more efficiently, which, obviously, has a positive impact on speed and performance regardless of system configuration.

The end product application was thus flexible, fast, and provided high-quality data capture in most respects. The remaining issue was to evaluate was medication coverage. As such, we compared the list of medications derived from RxNorm for this application to the existent legacy EHR medication data in Centerstone's system over the course of 12 months (July 2010 – June 2011). The legacy EHR medication data was populated via Infoscriber (http://ntst.com/products/infoscriber.asp), which captures standard medication names in a structured format. A very simple matching algorithm that performed a straight match on medication name (i.e., where EHR Medication Name = RxNorm Medication Name, accounting for capitalization differences) was able to correctly cover 93.2% (85,795/92,058) of all prescriptions during that year-long time period. Many of those non-matching medications were due to slight variations in naming (such as for "extended release versions of some medications, e.g., Venlafaxine vs. Venlafaxine ER, Seroquel vs. Seroquel XR) that could be easily correctly using a more complicated matching algorithm such as one based on regular expression pattern matching [4,12,18]. The other non-matching medications tended to be uncommonly prescribed medications and/or alternative therapy options such as nicotine cessation products, multivitamins, and dietary supplements (e.g., Omega-3 Fatty Acids). These non-matching patterns are similar to findings reported elsewhere [19]. The overall coverage was thus judged to be adequate for the intended purpose in this case – capturing medication history that might impact current treatment decisions and/or the evaluation of current treatment effectiveness – given that 1) the coverage rates of prescription orders were in line with previously reported values [7,9,20] and, 2) the non-matching prescriptions were generally of a nature (e.g., spelling variations, vitamins) that would not interfere with that intended purpose.

**4. Discussion**

As evidenced by the work presented here, RxNorm is a suitable terminology for capturing medication history in live EHRs. A restructured dataset of RxNorm satisfies the necessary requirements. It provides fast and efficient data access for high I/O applications and Google-like auto-complete search functionality patterned off of the Medicare Part D Formulary Finder website (http://plancompare.medicare.gov/pfdn/FormularyFinder/). It allows for flexibility to capture data in various, alternate clinical scenarios, such as when some data points are unknown or missing and RxNorm must be reconstructed from available component parts. Even when providing such flexibility, the restructured RxNorm dataset allows capture of data with reasonable data integrity and quality. A restructured RxNorm dataset can also provide excellent medication coverage, estimated at 93.2% in a large-scale, real-world outpatient behavioral healthcare setting. Many of the prescriptions not covered were nicotine cessation products, multivitamins, and dietary supplements or due to slight spelling variations of medication name.

A number of other research efforts have obtained similar results in similar projects related to RxNorm, as well as other standardized terminologies such as SNOMED CT [4]. Fung et al. [5] extracted a subset of commonly prescribed medications into a tool called "RxTerms". Zeng et al. [8] developed a PHR-like tool called "MyMedicationList" that partially uses RxNorm for direct capture of medication history directly from patients, which was later extended to interact with provider/patient collaborate e-prescribing tool called "MyRxPad" [21]. Nelson et al. [1] summarized many of these efforts. Other researchers have focused efforts on utilizing RxNorm as part of specific research projects [2,19]. All of the aforementioned efforts have produced insights regarding application of RxNorm toward developing web-based tools, personal health records, and specific research studies. This paper extends these studies by exploring the necessary steps toward implementation of RxNorm in an adapted form in EHRs. While many of the issues described elsewhere were similar to those presented here – e.g., coverage, flexibility, speed – some of the requirements for such an EHR application proved to be distinct, particularly in regards to I/O demands to support advanced caching and search functionality as well as flexible data capture across potential clinical scenarios. Future applications of RxNorm in EHRs in other live clinical settings may further elucidate these issues.

There are multiple potential avenues for extension of this work, particularly in regard to search optimization and filtering as well as integrating drug classification schemes into RxNorm.  The search results could be improved via the use of web-search-like indexing and filtering, similar to commonly used information retrieval techniques utilizing click-throughs and page ranking [22].  Such algorithmic approaches to search optimization can also incorporate user profiling to personalize medication search results based on individual user behavior or behavior of users similar to the current individual [23].  In other words, the search results that a particular user (patient or provider) sees when entering medication information may be ranked or organized based on other medications they have entered during the current session or based on past user behavior.  For example, medications that are commonly prescribed in "clusters" for co-occurring disorders may be pushed to the top of the list if the patient is identified as fitting that profile or if one medicine from such a cluster is entered.  Additionally, search functionality could be enhanced through the use of "sounds like" technology, such as Metaphone, that allows users to phonetically spell medication names and find them via search even if they do not know the correct spelling [22].  On other fronts, efforts to classify RxNorm data into medication categories hold great promise for research, clinical decision support, and business intelligence purposes, whereas RxNorm currently lacks any such built-in categorization scheme..  Grouping medications into categories, such as by linking to NDF-RT, enables cross-organizational, cross-population, cross-provider, and cross-diagnostic comparisons, among other capabilities.  However, there is still progress to be made in this arena, particularly in achieving better coverage/linkage of medications between RxNorm and drug categorization schemes [7,19].

**Acknowledgements**

This commentary and related research is funded by the Ayers Foundation and the Joe C. Davis Foundation. The funders had no role in the design, implementation, or analysis of this research. The author would also like to recognize various Centerstone staff for their contributions to this effort: Stephen Terrell and Prasad Kodali for their technical assistance, Dr. Frank Stevens and Matt Holfelner for human factors/usability testing, and Dr. Dennis Morrison, Dr. Tom Doub, and Christina Van Regenmorter for their efforts around implementation. The author would also like to thank Dr. Kay Connelly and Dr. Kelly Caine from Indiana University, as well as Dr. Rick Shelton from Vanderbilt University, for their various assistance in this work. The opinions expressed herein do not necessarily reflect the views of Centerstone or its affiliates.